# Generation of squeezed vacuum state in the millihertz frequency band

**Li Gao[1,3], Li-ang Zheng[1,3], Bo Lu[1], Shaoping Shi[1,*], Long Tian[1,2], and Yaohui Zheng[1,2,**]**

[1]State Key Laboratory of Quantum Optics and Quantum Optics Devices, Institute of Opto-Electronics, Shanxi University, Taiyuan 030006, China.

[2]Collaborative Innovation Center of Extreme Optics, Shanxi University, Taiyuan 030006, China.

[3]These authors contributed equally to this work.

*ssp4208@sxu.edu.cn
**yhzheng@sxu.edu.cn

**Abstract：**

The detection of gravitational waves has ushered in a new era of observing the universe. Quantum resource advantages offer significant enhancements to the sensitivity of gravitational wave observatories. While squeezed states for ground-based gravitational wave detection have received marked attention, the generation of squeezed states suitable for mid-to-low-frequency detection has remained unexplored. To address the gap in squeezed state optical fields at ultra-low frequencies, we report on the first direct observation of a squeezed vacuum field until Fourier frequency of 4 millihertz with the quantum noise reduction of up to 8 dB, by the employment of a multiple noise suppression scheme. Our work provides quantum resources for future gravitational wave observatories, facilitating the development of quantum precision measurement.

**Introduction**

The first direct detection of gravitational waves (GWs) opened a new chapter in astronomy, which announced the beginning of the era of gravitational wave astronomy [1]-[2]. Following the application of squeezed vacuum states of light, as well as other technological advances[3]-[4], all of the ground-based gravitational wave observatories, including Advanced Virgo[5], Advanced LIGO[6], KAGRA[7], and GEO600[8] have been operating with a quantum enhanced sensitivity[9]-[14], extending their astrophysical reach. In the long run, third-generation detectors, including the Einstein Telescope[15], and Cosmic Explorer[16], are expected to drive the audio-band reach out to cosmological distances. Space-borne laser interferometers would give us exquisitely sensitive probes for millihertz (mHz) band astrophysical signals[17]-[18]. Nonetheless, gaps still exist between ground-based and space-borne laser interferometers, which limits our ability to understand the elusive intermediate-mass black holes[19], corresponding to around deci-Hertz band[20]. Therefore, some conceptual designs of GW observatories have been proposed to fill in the gap in the 0.1-10 Hz band, such as a lunar-based interferometer, GLOC[21], and a ground-based interferometer, MANGO[22].

Following these conceptual designs that cover from 0.1 Hz to 10 Hz, one of the key issues that need to be addressed is the squeezed state generation in matching frequency bands with these conceptual GW observatories, thus meeting the needs of the GW community for the foreseeable future. Since the squeezed state was demonstrated experimentally in 1985[23], intense research activities for high squeezing factor[24]-[27], suitable squeezing frequency bands[28]-[29], stable phase control[30]-[32], and diverse operating wavelengths[33] are ongoing through various approaches. Four-wave mixing process can directly generate the squeezed state that matches with the atomic transition line and is almost perfectly adapted to the quantum repeater. In virtue of the two-beam noise-cancellation technique, the squeezing frequency band has been extended to below 1 Hz[34]-[35]. Optical parametric oscillator (OPO) is another scheme of squeezed state generation that continues to hold the record of the squeezing strength, the maximum squeezing factor has reached the value of 15 dB at l064 nm without active phase control[26]. Active control of the squeezed quadrature angle is the basic requirement for squeezed state applications. Unfortunately, the squeezed vacuum state has no coherent amplitude, making the extraction of error signals of active control impossible. By introducing a faint, phase-modulated seed field at the carrier frequency into the OPO, the issue about the active control is addressed, but inducing inevitable nonlinear noise coupling[36], as well as the transfer of seed laser technical noise that is considerably high at low-frequency[37]. In virtue of a coherent but frequency shifted beam, active control can be perfectly implemented without the seed laser injected[30]. The auxiliary control scheme which is compatible with the downstream application is immune to the low-frequency noise to some extent, gradually pushing the squeezing frequency band to the Hz level in combination with the intensity noise suppression of the pump beam[38]-[40]. Currently, the lowest squeezing frequency band generated by OPO is 0.5 Hz with 3 dB quantum noise reduction[41], covering the detection frequency band of ground-based GW observatories, but not meeting the requirement for GW observatories under conceptual design. Attributed to intrinsic thermal noise and detection noise etc[42], squeezed state generation at below Hz band is still an elusive task.

In this paper, we report on the first direct observation of a squeezed vacuum field based on OPO until Fourier frequency of 4 mHz with a quantum noise reduction of up to 8 dB. By the employment of a multiple noise suppression scheme proposed by us, all noise sources that restrict the low-frequency performance, including laser pointing noise, work-point drifting noise (coming from the displacement of the piezoelectric transducer), etc., are unified to laser intensity noise, further suppressing the unified intensity noise in virtue of a feedback loop. The obstacle that limits the squeezed state generation in the mHz-Hz band is removed. Incorporating the corresponding-band EOM (BHD), the squeezed state downwards to 4 mHz frequency band is directly detected. It covers the detection bands of the GW observatory under conceptual design, laying the foundation for future quantum precision measurements in the ultra-low frequency band.

**Results**

**Principle framework**

A simplified optical path of the squeezed vacuum state generation is shown in Fig. 1a. It can be seen from the arrow direction that the two beams injecting the OPO in opposite directions are the auxiliary control beam ($\omega+\Omega$) and pump beam ($2\omega$) respectively. The pump field is used for parametric down conversion to produce squeezed light ($\omega$), meanwhile, the auxiliary field also participates in the nonlinear interaction to control the squeezed angle via the generated two sidebands ($\omega+\Omega$, $\omega-\Omega$). The $2\Omega$ signal generated by the sideband beat frequency enters the servo system and feeds back to the phase shifter (PS, high-reflectivity mirror mounted on a piezoelectric transducer) placed in the pump beam to control the squeezed angle. However, any environmental disturbance may cause the phase shift of the squeezed angle (i.e. squeezed angle rotation), a PS located at the optical path of the pump beam is actively controlled to provide a tiny displacement, aiming to compensate for the phase shift. Unfortunately, the tiny displacement inevitably induces the beam pointing noise which is called work-point drifting noise. Considering a Gaussian pump beam is injected into the OPO, the mode matching efficiency depends on two error terms, including a mismatched parameter between the optical mode and the cavity mode, and a misaligned parameter coming from the deviation of the optical axis direction[43]-[44]. Similar to laser pointing noise, the work-point drifting noise changes the mode matching efficiency between the pump beam and OPO (more details are provided in Supplementary Material), as well as the injected intra-cavity pump power. The weak fluctuation is enough to induce the variation of the central temperature of the periodically poled KTiOPO₄ (PPKTP) crystal, further destroying the co-resonance condition[45]. As a result, the low-frequency squeezing noise degrades. The logic cycle of system fluctuation is in detail shown in Fig. 1b and Fig. 1c.

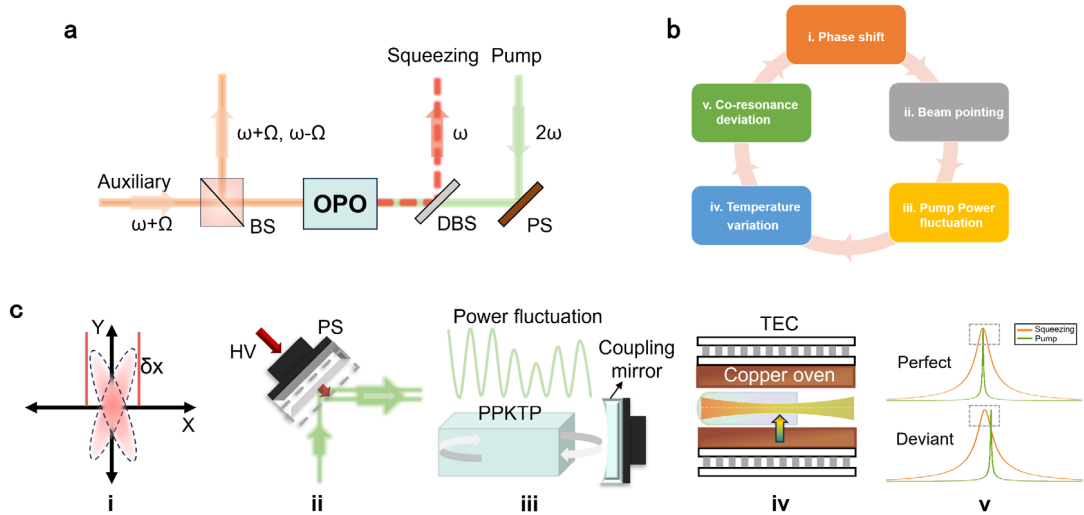

**Fig. 1 Evolution process from system drift to squeezing factor fluctuation. a** Schematic of the squeezed vacuum field preparation. BS: Beam splitter, DBS: Dichroic beam splitter, OPO: Optical parametric oscillator, PS: Phase shifter, ω: Squeezing light, 2ω: Pump beam, ω+Ω, ω-Ω: Sideband fields. The reflected signal from BS is extracted and fed back to PS to lock the squeezed angle. **b** Logic cycle of system fluctuations. **c** Sketch map of the evolution process. HV: High voltage, TEC: Thermoelectric cooler. The δx represents the shift of the phase lock point as seen in inset i "Phase shift". In inset ii "Beam pointing", the active control means adjusting the position of the PS to calibrate the squeezed angle. The dashed gray line is the position of the mirror after displacement, and the green line represents the pump beam before and after the compensation, indicating a tiny translation. It should be noted in inset iii "Power fluctuation" that the green curve represents the intra-cavity pump power fluctuation due to pump beam pointing noise. The crystal oven structure of the PPKTP crystal is shown in inset iv "Temperature variation", where the crystal is central heating by the absorbed pump laser and boundary cooling by a copper oven. The above and below figures in inset v "co-resonance deviation" indicate the co-resonance status without and with the phase shift, respectively, which depends on the central temperature of the crystal.

Here, taking a co-resonance OPO applied in our experiment as the example[46]-[47], we introduce the origin of the squeezing factor variation in the low-frequency band. The simultaneous resonance of the fundamental and pump beam is tuned via the stabilized crystal temperature, and the cavity length is held on resonance relying on the detection of the pump beam. Following the variation of mode matching efficiency, as shown in Fig. 1c(iii), the circulating power in the OPO fluctuates with time. In terms of the temperature control means shown in Fig. 1c(iv), there exists a thermal gradient across the cross-section of the PPKTP crystal. Although the boundary temperature is precisely stabilized, the central temperature inevitably fluctuates with the circulation power cross-section (more details are provided in Supplementary Material). Thus, the pre-optimization co-resonance status in the initial phase may deviate from the optimal point over time. This restriction often leads to a low level of the low-frequency squeezing factor, which is usually necessitating prolonged data acquisition. Existing demonstrations combine a coherent control scheme[30] and intensity noise suppression of the pump beam, driving the squeezing frequency band to 0.5 Hz[41]. For lower frequency, it is necessary to present an innovative experimental scheme.

**Experimental setup**

A schematic of the experimental setup is shown in Fig. 2. The laser source is a non-planar ring oscillator (NPRO, from Coherent Inc.) laser with 2.0 W continuous-wave single-frequency output power at 1064 nm. The laser preparations, including spatial-mode improvement, polarization purification, and laser power stabilization are similar to our earlier experiments presented in Refs. [48]-[49]. The design of the second harmonic generation (SHG) cavity is similar to the OPO cavity described below. The OPO is a hemilithic optical resonator consisting of a PPKTP crystal. A copper oven is designed according to the size of the crystal, with the crystal wrapped in indium foil.

The temperature of the crystal is controlled by a thermoelectric cooler, resulting in a relatively stable temperature field distribution inside the crystal. By finely adjusting the temperature control, the fundamental field and the pump field can be co-resonant in the OPO.

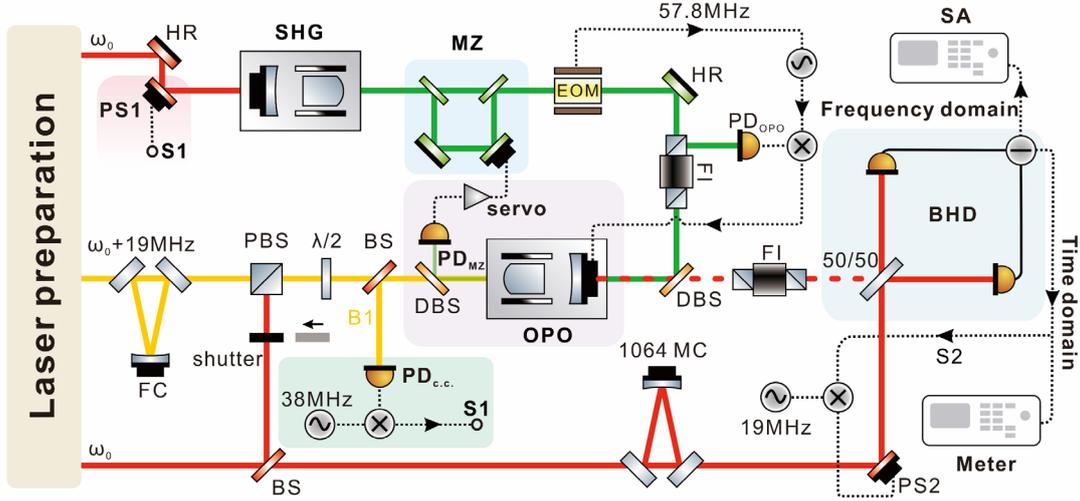

**Fig. 2 Experimental setup of squeezed state generation and detection at mHz frequencies.** PS: Phase shifter, HR: High reflectivity mirror, SHG: Second harmonic generation, EOM: Electro-optic modulator, MZ: Mach-Zehnder interferometer, $\lambda/2$: Half-wave plate, FI: Faraday isolator, PBS: Polarizing beam splitter, BS: Beam splitter, DBS: Dichroic beam splitter, OPO: Optical parametric oscillator, FC: Filter cavity, MC: Mode cleaner, S1: Error signal 1, S2: Error signal 2, PD: Photoelectric detector, BHD: Balanced homodyne detector, SA: Spectrum analyzer, Meter: Multimeter.

The phase modulated signal with the modulation frequency of 57.8 MHz is imprinted on the pump beam by an electro-optic modulator to extract the error signal for controlling the OPO cavity length and further drives the generation of the squeezed vacuum field. Following the coherent control scheme[30], we employ a frequency-shifted auxiliary control field that is $\Delta$=19 MHz away from the main laser frequency $\omega_0$. In addition, the filter cavity located at the optical path of the auxiliary beam can efficiently suppress the undesired fundamental $\omega_0$ and improve the spatial-mode degradation caused by the acousto-optic effect. The auxiliary field has a weak nonlinear interaction with the pump beam, resulting in the generation of a sideband at a frequency of -19 MHz. The corresponding reflective field (B1) from OPO carries the phase information inside the cavity. The error signal S1 is obtained by demodulating the photocurrent at twice the beat frequency (38 MHz), feeding back to PS1 to control the squeezed angle.

We achieve the laser power pre-stabilization in the laser preparation section. However, as shown in the principle framework, the circulating power of the pump field inside the OPO will fluctuate with not only the pump beam intensity noise but also the work-point drifting noise caused by the displacement of PS. In virtue of an elegant scheme that places the PS (the position PS1 in Fig. 2) in front of the SHG, instead of behind the SHG, these noise sources that restrict the low-frequency performance, including laser pointing noise, work-point drifting noise are unified to laser intensity noise. A low-noise detector PDmz is located at the transmission end of the OPO to extract the error signal of the intensity noise feedback loop. Therefore, the pump power is not simply locked to a constant voltage reference but real-time compensation with the circulating power of the OPO. Compared with the common laser power stabilization scheme[39], our scheme can stabilize the circulating pump power of the OPO, rather than the input pump beam power. Ultimately, the long-term stability of the OPO cavity is achieved through the above multiple noise suppression scheme.

The 1064 MC is employed to improve the spatial mode and suppress high frequency noise. The output beam from the 50/50 beam splitter is directed toward a BHD to observe the noise level. During the measurement process, the seed field is blocked, thereby eliminating the nonlinear noise coupling. To detect the squeezed state at lower frequencies, we have developed an ultra-low noise and high-gain BHD in the mHz frequency band. The relative phase between the squeezed field and local oscillator (LO) is actively compensated with PS2 by utilizing the Pound-Drever-Hall technique, where the necessary error signal S2 is obtained from demodulating a part of the BHD AC signal with a 19 MHz sinusoidal signal. Further details regarding the experimental setup can be found in the Materials and Methods.

**Experimental results**

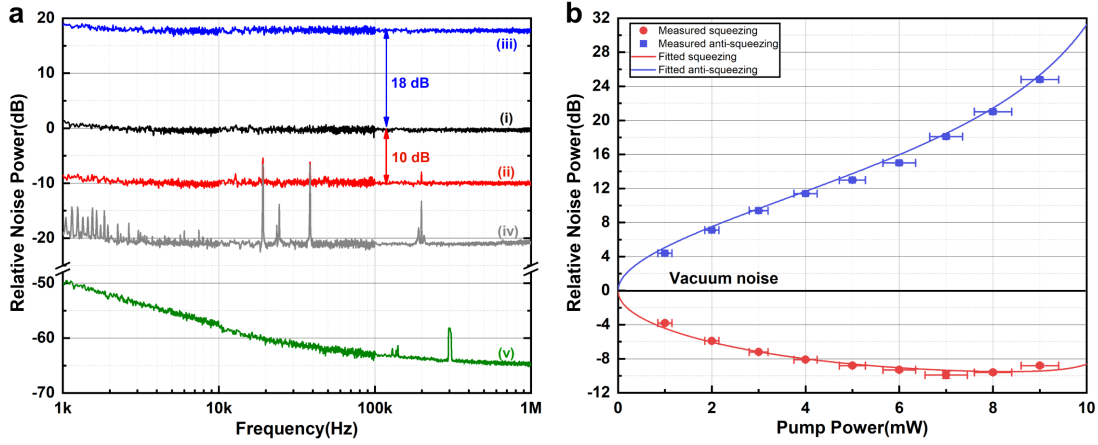

**Fig. 3 Quantum noise measurements and squeezing/anti-squeezing fitting curves.**
**a** Measured noise spectra from 1 kHz to 1 MHz. Trace (i)-(v) represent the shot noise, squeezing, anti-squeezing, electronic noise of BHD, and noise floor of spectrum analyzer (RS: FSW8), respectively. All traces are combined of three Fast Fourier Transform (FFT) frequency windows acquired with the spectrum analyzer: 1 kHz-10 kHz, 10 kHz-100 kHz, and 100 kHz-1 MHz, with resolution bandwidths of 20 Hz, 100 Hz, and 100 Hz, respectively. 200 root mean square averages are taken for all traces. **b** Pump power dependence of the squeezing and anti-squeezing spectra. All values are obtained from zero-span measurements at the analysis frequency of 1 MHz. All the data are still including electronic noise and normalized to the vacuum reference. The fitting curves, (the blue and red solid lines) are modeled using Eqs. (1) for the phase noise and loss characterization. The absolute error of a given pump power is 4.5% due to the measurement uncertainty of the pump meter. Measurements of squeezing and anti-squeezing factors are repeated more than 20 times at each power level, and then the standard deviations are calculated as the vertical error bars.

Fig. 3a shows the variances of squeezed states between 1 kHz-1 MHz. Trace (i) represents the homodyne detector shot noise reference measured with the LO beam of 2.8 mW intensity. With the pump power of 7 mW, an almost smooth quantum noise reduction of at least 10.0 dB from 2 kHz to 1MHz can be directly measured (trace(ii)) without electronic noise subtraction, whereas the corresponding anti-squeezing is amplified by 18.0 dB (trace (iii)). Trace (iv) corresponds to the electronic noise of BHD and is recorded with the signal and LO input blocked. The electronic noise of BHD is about 21.0 dB below the shot noise. Trace (v) corresponds to the noise floor of the measuring instrument. The uptick in the noise spectrum at around 1kHz is believed to be caused by the coupling of BHD electronics noise.

For characterizing the phase noise and losses in the experimental setup, we compared the measurement results with a theoretical model. The squeezed and anti-squeezed quadrature variances for an ideal OPO below the threshold can be computed as[49]:

$$V_{s/a} = [1 \mp \eta_{tot} \frac{4\sqrt{P/P_{thr}}}{(1 \pm \sqrt{P/P_{thr}})^2 + 4(2\pi f/\gamma)^2}]\cos^2 \theta$$
$$+[1 \pm \eta_{tot} \frac{4\sqrt{P/P_{thr}}}{(1 \mp \sqrt{P/P_{thr}})^2 + 4(2\pi f/\gamma)^2}]\sin^2 \theta$$

where $P$ is the pump power, $P_{thr}$ is the threshold power, $\theta$ is the phase fluctuation, $\eta_{tot}$ is the total detection efficiency, and $f$ is the Fourier frequency. The cavity decay rate $\gamma$ can be calculated as $\gamma = c(T+L)/2l$, with the vacuum speed of light $c$, the cavity single-trip length $l$, the power transmissivity of coupling mirror $T$, and the total intracavity linear loss $L$. The squeezing and anti-squeezing level dependences of pump power are shown in Fig. 3b. Fitting our measurement results with the model, we can obtain that the total efficiency $\eta_{tot} = 90.3\%$ and the total phase fluctuation $\theta = 5$ mrad. It should be noted that the actual threshold of the co-resonance OPO varies with the pump power injected into the cavity. This variability contributes to discrepancies between measured values and the fitted curve.

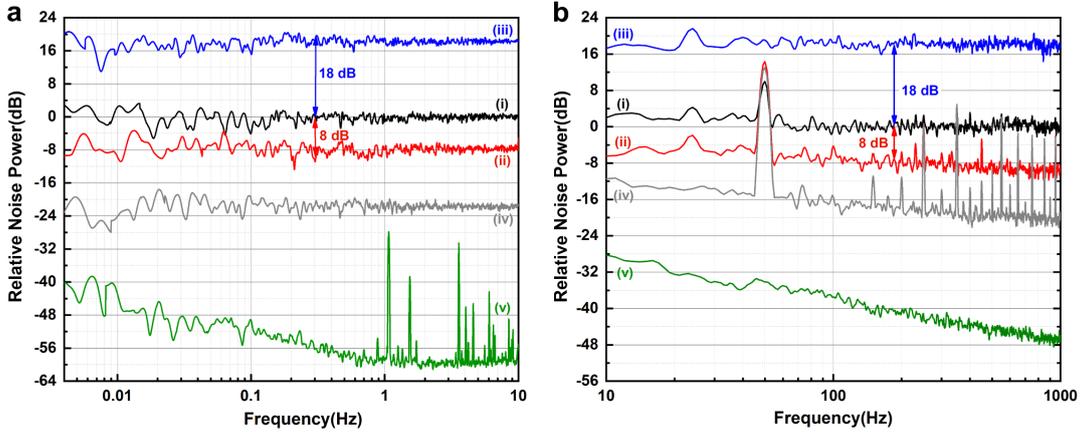

**Fig. 4 Quantum noise measurements from 1 kHz downwards to 4 mHz. a** Acquire from high-precision multimeter (Keithley: 3706A), with the sampling frequency of 20 Hz. **b** Acquire from spectrum analyzer (RS: FSW8), with 3 Hz resolution bandwidth and 1 Hz video bandwidth. In both **a** and **b**, Trace (i) represents the shot noise, Trace (ii) is squeezing noise, Trace (iii) is anti-squeezing noise, Trace (iv) is BHD electronic noise, and Trace (v) is measurement instrument (3706A/ FSW8) noise.

In this configuration, we perform squeezing noise measurements below 1 kHz by utilizing the multiple noise suppression scheme. With the same measurement conditions for traces (i-iii) in Fig. 3, we achieve the normalized measured results of the noise power are shown in Fig. 4. It is important to note that Fig. 4a and Fig. 4b are measured using different instruments to cover the respective frequency bands, as a result of sampling precision limitations. In Fig. 4a, we utilize an accurate digital multimeter to sample the data in the time domain, then employ a Logarithmic power spectrum density (LPSD) algorithm to perform the Fourier transform. Each trace is computed from a 40-min-long time series, sampled with 20 S/s. To verify the accuracy of our algorithm and ensure that the obtained results are convincing, the noise power spectrum is directly measured by the spectrum analyzer from 10 Hz to 1 kHz, shown in Fig. 4b. The obtained squeezing level can be correlated with the data below 10 Hz. We suspect that the non-flat quantum noise and dark noise observed in Fig. 4b are caused by the rapid increase in noise floor of the spectrum analyzer as the measurement frequency decreases.

A quantum noise reduction of up to 8.0 dB from 300 Hz downwards to 4 mHz could be fully displayed (trace(ii) of Fig.4a and Fig. 4b). Benefit from the excellent clearance of home-made BHD and the extremely low noise floor of the multimeter Keithley 3706A, we also obtain a flat squeezing noise spectrum at 0.3 Hz-10 Hz. The noise peak around 50 Hz in Fig. 4b comes from the power supply of the BHD. At the analysis frequency of 1 kHz, it still maintains the squeezing level near 10.0 dB. Compared to the experimental results within the 1 kHz-1 MHz frequency range shown in Fig. 3a, the squeezing level below 1 kHz has been slowly degraded with the analysis frequency. This may result from several increased noise couplings at the low measurement frequencies[42]. For example, the scattering loss due to air flow or particles and the residual technical noise, etc. show up at the unbalanced homodyne detector, leading to the squeezing degradation.

According to known experimental parameters, we can directly obtain the escape efficiency of the OPO[50], the interference visibility, and the propagation efficiency. The interference visibility is 99.0±0.1%, corresponding to an optical loss of 2.0%. By the employment of the total loss fitted above, the quantum efficiency of the photodiode can be inferred. The loss budget for the squeezed system is listed in Table 1.

The propagation loss mainly comes from the Faraday isolator with an insertion loss of 3.5%, as well as non-perfectly coated mirrors with 1.1% of residual reflection and transmission losses. By employing a ring cavity as the OPO, parasitic interference can be significantly suppressed without the use of the Faraday isolator[41]. In addition, the BHD is solely meant to measure the squeezing factor, which has no use for practical downstream applications. Therefore, we should deduct the contribution of the detection loss, to infer the squeezing factor that can be applied in the downstream experiment. Without considering the insertion loss of the Faraday isolator, 10.5 dB of squeezed state in the mHz frequency band is expected to apply to a practical experiment system. The comprehensive noise-suppression scheme, integrated with the advanced polishing and coating method, will make the highest squeezing factor in mHz frequency tangible.

**Table 1 Loss budget for the squeezed state generation system**

| Source of Loss | Value |
|---|---|
| OPO escape efficiency | 98±0.45% |
| Propagation efficiency | 95.4±0.2% |
| 99% homodyne visibility | $(99.0\pm0.1\%)^2$ |
| Photodiode quantum efficiency | 98.5±0.2% |
| Total efficiency | $90.3\pm0.15\%$ |

**Discussion**

A detailed analysis is conducted on the work-point drifting noise that originates from inevitable environmental disturbance. We propose a multiple noise suppression scheme, addressing the obstacle of restricting the low-frequency performance of the squeezed state. Exploiting the SHG cavity as a noise-conversion element, laser pointing noise and work-point drifting noise are unified to laser intensity noise. In combination with a feedback loop, the unified intensity noise which in turn induces the temperature fluctuation of the PPKTP crystal and the co-resonance status deviation is suppressed.

In conclusion, we have unlocked the generation and detection of squeezed state in the mHz frequency band, which is two orders of magnitude lower than the currently reported lowest frequency, and it meets the requirements for future GW observatory detection in 0.1-10 Hz frequency bands. At the same time, the squeezing factor detected reaches 8.0 dB at the mHz level. Without regard to the detection loss, 10.5 dB of the squeezed state is expected to apply to a practical GW observatory in the mHz frequency band. This work is prepared for the detection of faint astronomical signals in future mid-frequency and even lower frequency ranges.

**Materials and methods**

Approximately 100 mW of fundamental laser is injected into the SHG cavity, resulting in a 60% conversion efficiency. About 7 mW of 532 nm pump light from SHG is sent to the OPO, resulting in the pump factor of 0.7 and the classical gain of 20. Different from our previous experimental scheme, the absence of the pump light mode cleaner can shorten the optical path and reduce the influence of environmental disturbance.

The OPO is a hemilithic optical resonator consisting of a PPKTP crystal with the dimensions $1.0 \times 2.0 \times 10.0$ mm$^3$ and a piezo mounted output coupling mirror. For co-resonance, we have chosen the power reflectivity of the coupling mirror with 25 mm radius of curvature is 85% at 1064 nm and 97.5% at 532 nm. The back surface of the crystal with 12 mm radius of curvature has a high reflection coating (R>99.9%) for both wavelengths. The cavity's free spectral range and bandwidth of 1064 nm (532 nm) are 3.83 GHz (3.76 GHz) and 98.8 MHz (15.8 MHz), respectively. Depending on the pump power injected into the OPO, the temperature of the chamber is carefully controlled at 32 °C.

The Faraday isolator can effectively reduce the influence of parasitic interference on low-frequency squeezing noise. Thus, the squeezed light emitted from the concave mirror of OPO is separated from the pump light by a DBS, passed through the Faraday isolator with 96.5% transmissivity, and carefully mode matched with the LO on a 50/50 beam splitter. It is worth noting that the lower frequency bands measured, the greater the influence of classical noise. This requires not only the BHD with a high common-mode rejection ratio but also an accurate power balance that is coupled into two photodiodes.

Electronic noise, also known as dark noise, is another important factor limiting the measurement of low-frequency shot noise. We use a two-stage amplifier circuit so that the AC of the BHD has enough gain. After repeated comparison, the best low-noise components are selected for the amplifier circuit. The photocurrents of both photodiodes are subtracted before entering the amplifier circuit to further reduce the imbalance due to the component differences. In the lower frequency band, another non-negligible electronic noise is flicker noise. We choose metal film-type resistors to reduce the impact of flicker noise on the BHD[42]. A low noise and high gain BHD ensures that the squeezed vacuum state at mHz frequencies is effectively detected.

**Acknowledgments**

The project is sponsored by National Natural Science Foundation of China (NSFC) (Nos. 62225504, 62027821, U22A6003, 62035015, 12174234, 12304399); National Key R&D Program of China (Grant No. 2020YFC2200402); Fundamental Research Program of Shanxi Province (No.202303021212003, 202303021224006).

**Author Contributions**

L.G., L.-A.Z. and S.-P.S design the experiment. L.G. and L.-A.Z. carry out the experiment with assistance from S.-P.S and L.T. B.L. and L.T. help collect the data. L.G., S.-P.S and Y.-H.Z. analyze the data and write the paper with input from all other authors. The project is supervised by Y.-H.Z. All authors discuss the experimental procedures and results.

**Data availability**

The authors declare that all data supporting the findings of this study can be found within the paper and its Supplementary information files. Additional data supporting the findings of this study are available from the corresponding author (Y. H. Z.) upon reasonable request.

**Conflict of interest**

The authors declare no competing interests.

**Supplementary information**

See the supplementary information for additional results, supporting figures, and data analysis procedures.

(2018).